\documentclass[twocolumn,twoside]{IEEEtran}

\ifCLASSINFOpdf

\else

\fi

\ifCLASSOPTIONcompsoc
\usepackage[caption=false, font=normalsize, labelfont=sf, textfont=sf]{subfig}
\else
\usepackage[caption=false, font=normalsize]{subfig}
\fi
\usepackage{lipsum}%
\usepackage[dvipsnames]{xcolor}

\usepackage{balance}
\usepackage{multicol}   
\usepackage{cite}
\usepackage{gensymb}
\usepackage{multirow}
\usepackage{graphics}
\usepackage{epsfig}
\usepackage{graphicx}
\usepackage{epstopdf}
\usepackage{textcomp}
\usepackage{amsmath}
\usepackage{mathtools}
\usepackage{filecontents}
\usepackage{lipsum,color}
\usepackage{amssymb}
\usepackage{float}
\usepackage{hyperref}
\usepackage{amsthm}

\usepackage{algorithm}
\usepackage{algpseudocode}

\usepackage{times} 
\usepackage{amsthm}  

\usepackage{amsfonts}

\theoremstyle{break}

\begin{document}

\title{Picosecond Wireless Synchronization with Entangled Photons via Grid-Based Quantum Coverage in Indoor Optical Systems}

\author{Hossein~Safi, ~{\it Member,~IEEE}, ~Mohammad~Taghi~Dabiri,~Mazen~Hasna,  ~{\it Senior Member,~IEEE}, ~Iman~Tavakkolnia,~{\it Senior Member,~IEEE},
	~and~Harald~Haas,~{\it Fellow,~IEEE}
	\thanks{H. Safi, I. Tavakkolnia, and H. Haas are with the Electrical Division, Department of Engineering, University of Cambridge, Cambridge, UK, e-mails: \{hs905, it360, huh21\}@cam.ac.uk.}
	\thanks{Mohammad Taghi Dabiri is with the College of Science and Engineering, Hamad Bin Khalifa University, Doha, Qatar (E-mail: mdabiri@hbku.edu.qa). M. Hasna is with the Department of Electrical Engineering, Qatar University, Doha, Qatar, e-mail: hasna@qu.edu.qa.}
	\thanks{This work was supported by the UK Engineering and Physical Sciences Research Council (EPSRC) grant EP/Y037243/1 for TITAN Telecoms Hub.}
	
}

\maketitle
\begin{abstract}
	\textcolor{black}{In this paper, we present a robust entanglement-assisted synchronization framework for indoor optical wireless systems that explicitly captures the coupling between spatial beam geometry and temporal synchronization accuracy. Unlike conventional approaches that treat beam steering and timing estimation independently, a unified spatio temporal model is developed that links user position uncertainty to the Cramér Rao lower bound of the synchronization error. The framework incorporates key physical impairments, including multipath dispersion, non Gaussian detector jitter, and spatially correlated localization errors. Through analytical modeling and extensive simulations, we show that the proposed system exhibits graceful performance degradation under heavy tailed positioning uncertainty and remains stable in the presence of multipath induced bias. Using realistic single photon detector parameters, the results indicate that synchronization accuracy below $10$ picoseconds can be maintained across a wide range of operating conditions. This level of precision provides a scalable foundation for quantum enabled indoor wireless networks.}
\end{abstract}


\IEEEpeerreviewmaketitle


\section{Introduction}
\label{intro}

The convergence of quantum information science and wireless communication is enabling a new class of systems that support secure communication, high precision sensing, and distributed quantum processing \cite{9749227,9928082}. Among the available physical platforms, optical wireless communication has emerged as a particularly attractive candidate for realizing quantum enhanced networking functionalities, including quantum key distribution, quantum sensing, and local area quantum networks \cite{9928082,11021496,hasan2023quantum,xu2024optical,10974735}. A fundamental requirement across all such applications is the ability to distribute and detect entangled or single photons with extremely high temporal precision, typically at the picosecond level \cite{liu2019energy}.

This requirement becomes especially critical in protocols such as quantum key distribution and entanglement swapping, where the identification of correlated detection events relies on narrow coincidence windows. Even small clock offsets or uncertainties in propagation delay can significantly degrade detection fidelity, increase error rates, and compromise quantum coherence \cite{ge2024reduction,ge2023analysis}. Consequently, time synchronization is not merely an implementation detail but a core physical requirement for reliable quantum communication.
\begin{table*}[!t]  
	\color{black}  
	\centering
	\caption{System-Level Comparison of Representative Synchronization Schemes}
	\label{tab:sync_comparison}
	\renewcommand{\arraystretch}{1.2}
	\setlength{\tabcolsep}{4pt}
	\small
	\begin{tabular}{|l|c|c|c|c|}
		\hline
		\textbf{Feature} 
		& \textbf{Classical} 
		& \textbf{Two-Way} 
		& \textbf{Single-Beam} 
		& \textbf{Proposed} \\
		& \textbf{Optical} 
		& \textbf{Quantum} 
		& \textbf{Quantum} 
		& \textbf{Grid-Quantum} \\
		\hline
		Spatial Coverage
		& No 
		& No 
		& No 
		& {Yes} \\
		\hline
		Mobility Support
		& Limited 
		& No 
		& Limited 
		& {Yes} \\
		\hline
		Sync Time
		& Fast 
		& Long 
		& Moderate 
		& {Short} \\
		\hline
		Robust to Sparsity 
		& No 
		& Moderate 
		& Low 
		& {Yes} \\
		\hline
		Dominant Noise
		& Thermal 
		& Fiber Jitter 
		& Shot Noise 
		& {Shot Noise/Multipath} \\
		\hline
		Sensitivity
		& Poor (High power required)
		& Medium 
		& Good 
		& {Good} \\
		\hline
		Link Model
		& Continuous 
		& Static P2P 
		& Continuous 
		& {Grid-Based} \\
		\hline
		Handover Support
		& Complex (Re-locking) 
		& No 
		& No 
		& {Native (Fast Re-sync)} \\
		\hline
	\end{tabular}
	
	\vspace{2mm}
	\footnotesize
	\textit{Note:} Sensitivity and synchronization times are indicative, corresponding to different operating regimes. Classical schemes assume continuous high-SNR detection; quantum and SPAD schemes operate in photon-counting regimes with distinct noise statistics.
\end{table*}
Motivated by this need, research on quantum clock synchronization has progressed along two principal directions. The first focuses on long haul fiber based quantum links, where weak quantum signals must coexist with classical traffic \cite{burenkov2023synchronization,hong2023quantum}. These systems rely on techniques such as wavelength division multiplexing, polarization control, and dispersion compensation, and have recently demonstrated high precision synchronization using energy time entangled photons \cite{spiess2023clock,shi2024quantum}. The second direction targets free space quantum links, including ground to satellite communication, where synchronization must contend with atmospheric turbulence, background radiation, and Doppler effects \cite{liao2017long,cao2020long,xiang2023quantum}. While both approaches have achieved impressive performance, they are designed for relatively static links and rely on assumptions that do not hold in indoor environments.

\textcolor{black}{Indoor optical wireless systems operate under conditions that differ fundamentally from long-haul or point-to-point optical links. They must support user mobility, frequent handovers, dense spatial reuse, and operation in multipath-rich environments with ambient optical interference. OWC is well suited to these conditions due to its large available bandwidth, strong spatial confinement, and compatibility with single-photon detection technologies. However, these same characteristics impose stringent requirements on synchronization, as timing accuracy must be maintained despite spatial uncertainty and dynamic link reconfiguration. Consequently, indoor quantum synchronization schemes must be inherently robust to user motion and beam misalignment, rather than relying on static or continuously aligned links.}

\textcolor{black}{Despite recent progress in entanglement assisted synchronization and optical time transfer \cite{li2023entanglement}, most existing formulations implicitly decouple spatial beam configuration from timing performance. Synchronization accuracy is typically analyzed under fixed link assumptions, without accounting for the effects of beam footprint, user position uncertainty, or spatial coverage geometry on photon detection statistics. As a result, the interaction between spatial design and achievable timing precision remains largely unexplored, even though it is central to practical indoor optical wireless deployments.}

\textcolor{black}{At the same time, significant progress has been made in classical indoor optical wireless architectures. In particular, the grid of beams (GoB) paradigm has emerged as an effective solution for providing full spatial coverage and supporting user mobility \cite{kazemi2025novel,soltani2023terabit,zeng2021vcsel,kazemi2022tb,safi20253d}. In a GoB system, a dense array of static, narrow beams partitions the coverage area into fixed cells, and user mobility is handled through digital handovers rather than mechanical beam steering. Although this architecture offers strong scalability and robustness, it is inherently quantum-agnostic. Each handover modifies the optical path length and invalidates prior timing calibration, while existing GoB implementations lack mechanisms to rapidly re-establish synchronization at the precision required for quantum applications under user mobility.}

\textcolor{black}{In this work, we address this gap by developing a unified spatio temporal synchronization framework in which spatial grid geometry and timing performance are analytically coupled. Unlike prior approaches that treat beam steering and synchronization independently, we show that user position uncertainty directly determines the number of detected photon pairs and therefore the fundamental synchronization accuracy. This relationship is captured through an explicit Cramér Rao lower bound (CRLB) formulation. Thus, it enables quantitative evaluation of how spatial design choices translate into timing precision.}
\textcolor{black}{To clarify the limitations of existing approaches and to position the proposed method within the broader landscape, Table~\ref{tab:sync_comparison} presents a system level comparison of representative synchronization schemes. The comparison highlights that existing classical and quantum approaches either lack spatial scalability or fail to support mobility and sparse photon detection, motivating the need for a grid based quantum synchronization framework.}

Building on this insight, we particularly propose an entanglement assisted synchronization framework specifically tailored to grid based indoor optical wireless systems. The proposed approach combines the spatial scalability of GoB architectures with the temporal precision offered by entangled photon pairs. A two stage synchronization algorithm is then introduced, consisting of coarse alignment using sparse bit pattern matching followed by fine synchronization through timestamp averaging over correlated detection events. In parallel, a comprehensive analytical model is developed to capture photon pair statistics, Gaussian beam propagation, user position uncertainty, and timing jitter introduced by both the source and detectors. To the best of our knowledge, this is the first work to present a complete entanglement assisted synchronization framework explicitly designed for grid based indoor optical wireless networks.

 \textcolor{black}{The main contributions of this paper are summarized as follows:}
	\textcolor{black}{
\begin{itemize} 
	\item We introduce a unified system architecture that integrates grid based optical delivery with entanglement assisted synchronization. This design establishes the fundamental coupling between spatial beam footprint and achievable timing variance. 
	\item Next, we develop an end to end analytical model that extends beyond ideal assumptions to incorporate multipath induced dispersion, non Gaussian positioning errors, and heavy tailed detector jitter. 
	\item Also, we design a practical two stage synchronization algorithm tailored for sparse single photon detection and robust to the timing biases and spatial misalignments inherent to indoor mobility. 
	\item Finally, we conduct a comprehensive performance evaluation demonstrating sub 10 ps accuracy and verifying system stability under realistic non Gaussian error distributions.
\end{itemize}}

This work therefore provides a foundational step toward scalable and high precision quantum networking in indoor environments, bridging quantum synchronization theory with practical optical wireless system design.

\textcolor{black}{The remainder of this paper is organized as follows. Section~\ref{sysmode-sec} presents the system model. Section~\ref{modeling-sec} describes the spatial modeling of photon reception. Section~\ref{synchro-sec} introduces the synchronization framework and the associated theoretical analysis. Section~\ref{sec:crlb} derives the CRLB and establishes fundamental limits on synchronization accuracy. Section~\ref{sec:channel_impairments} introduces the channel impairments considered in the proposed framework. Simulation results and performance evaluation are presented in Section~\ref{simulations-sec}. Finally, Section~\ref{conclusion-seccc} concludes the paper and outlines directions for future research.}

%


\begin{table}[!h]
	\label{notations}
	\centering
	\caption{List of Key Notations}
	\label{tab:notations}
	\begin{tabular}{ll}
		\hline
		\multicolumn{2}{c}{\textbf{Physical Environment}} \\
		\hline
		$x, y, z$ & Global coordinate system \\
		$x', y', z'$ & Beam-aligned local coordinate system \\
		$L, W, H$ & Room length, width, and height \\
		$h_{\text{cov}}$ & Optical coverage height from the floor \\
		$N_x, N_y$ & Number of grids along $x$ and $y$ axes \\
		$N_g = N_x \times N_y$ & Total number of spatial grid zones \\
		$\mathbf{p}_{\text{tx}}$ & Transmitter location (origin) \\
		$\mathbf{p}_u$ & True user position in global coordinates \\
		$\hat{\mathbf{p}}_u$ & Estimated user position \\
		$\sigma_p^2$ & Position estimation error variance \\
		$\mathbf{d}^{(i,j)}$ & Beam direction vector toward grid $(i,j)$ \\
		$\mathbf{R}^{(i,j)}$ & Rotation matrix to beam-aligned coordinates \\
		$\mathbf{p}_u'$ & User position in local (beam) frame \\
		$r_a$ & Receiver aperture radius \\
		$\tau_\text{rms}$ & Channel delay spread \\
		\hline
		\multicolumn{2}{c}{\textbf{Transmitter Parameters}} \\
		\hline
		$\mu_t$ & Mean photon-pair generation rate per slot \\
		$\tau_p$ & Pump pulse width \\
		$t_g[k]$ & Photon-pair generation time in slot $k$ \\
		$\sigma_t$ & SPDC source temporal jitter \\
		$t_r[k]$ & Reference photon detection time at transmitter \\
		$n_r[k]$ & Jitter at transmitter SPAD (reference) \\
		$\sigma_d$ & SPAD detector timing jitter \\
		$\lambda$ & Photon wavelength \\
		$w_0$ & Beam waist at source \\
		$w(z)$ & Beam width at distance $z$ \\
		\hline
		\multicolumn{2}{c}{\textbf{Receiver Parameters}} \\
		\hline
		$P_{\text{recept}}^{(i,j)}$ & Photon reception probability in grid $(i,j)$ \\
		$\eta_d$ & SPAD quantum detection efficiency \\
		$\mu_s^{(i,j)}$ & Mean number of detected signal photons per slot \\
		$\mu_b$ & Mean number of background photons per slot \\
		$N_s[k]$ & Number of received signal photons in slot $k$ \\
		$N_b[k]$ & Number of background photons in slot $k$ \\
		$t_u[k]$ & Photon detection timestamp at user \\
		$n_u[k]$ & Receiver SPAD timing jitter \\
		$\tau_{\text{sync}}$ & Effective synchronization offset  \\
		\hline
	\end{tabular}
\end{table}

\section{System Model}
\label{sysmode-sec}
We consider an indoor quantum synchronization system based on a grid-structured optical wireless delivery architecture that utilizes polarization entangled photon pairs. A central quantum unit, mounted on the ceiling, contains a spontaneous parametric down-conversion (SPDC) source that continuously generates entangled photons. In each pair, one photon remains at the central unit as a precise timing reference, while the other is directed toward the user through one of $N_g$ spatially distinct optical wireless transmission paths. Each transmission path corresponds to a fixed illumination region (grid) on the indoor surface, effectively dividing the room into $N_g$ non-overlapping zones. In the following subsections, we provide a detailed description of the system model and the parameters considered in this paper. Meanwhile, a summary of the notations used is presented in Table \ref{notations}.

\begin{figure*}
	\begin{center}
		\includegraphics[width=5.6 in]{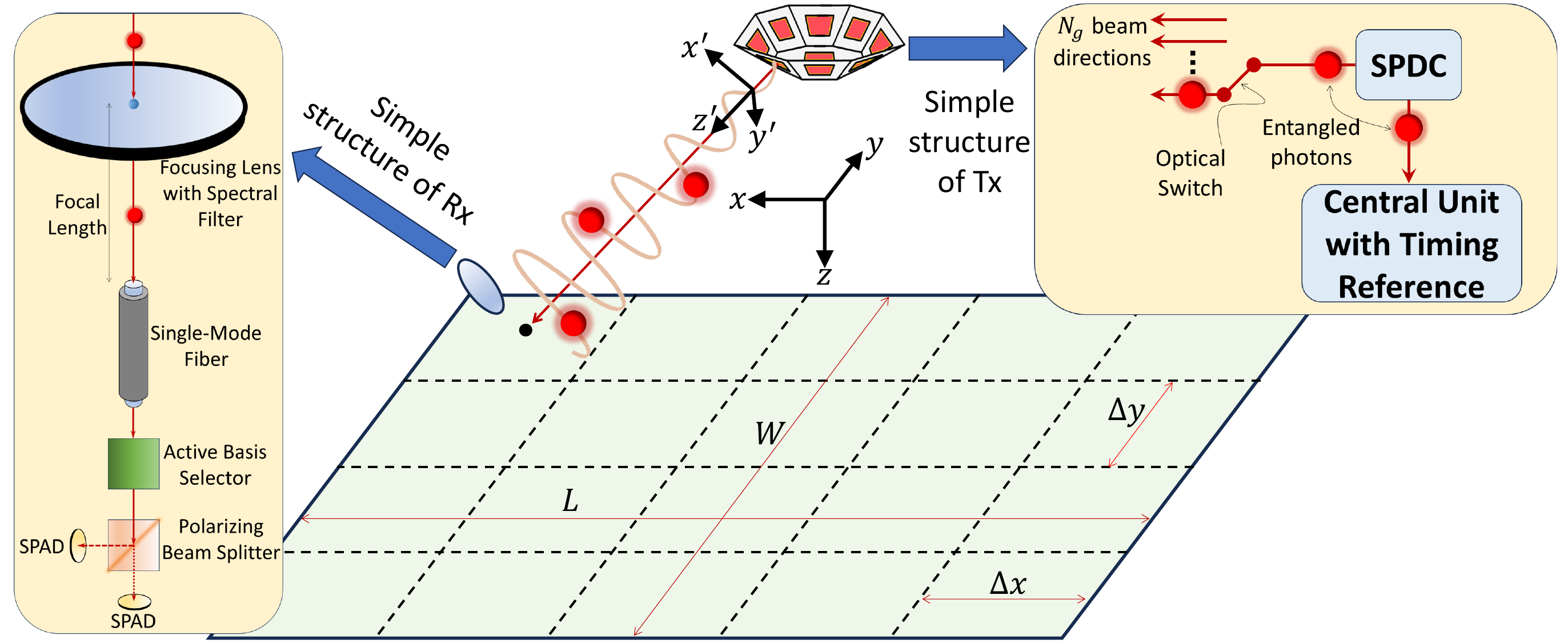}
		\caption{	System architecture of the proposed entanglement-based indoor quantum synchronization scheme. A central ceiling-mounted transmitter emits polarization encoded entangled photon pairs, directing the user photon toward a spatially selected beam corresponding to the estimated user location. The beam is aligned with a specific grid region in the room, while the reference photon is time-stamped at the transmitter. The receiver structure includes a circular aperture and SPAD array aligned to detect the photon and record its arrival time for synchronization. Local $(x', y', z')$ and global $(x, y, z)$ coordinate systems are depicted for spatial modeling.
		}
		\label{sm1}
	\end{center}
\end{figure*}

\subsection{Geometric Configuration and Coordinate Framework}
We consider a rectangular indoor environment with length $L$, width $W$, and height $H$. The quantum transmitter unit is mounted at the center of the ceiling, which is chosen as the origin of the global 3D coordinate system. Throughout this work, we denote global positions using the coordinate system $(x, y, z)$, with the transmitter located at $\mathbf{p}_{\text{tx}} = [0, 0, 0]$.
The room is partitioned into $N_g = N_x \times N_y$ non-overlapping grid zones, where $N_x$ and $N_y$ are the number of divisions along the $x$ and $y$ axes, respectively. Each grid is centered at a position $\mathbf{p}_{g}^{(i,j)}$ on the horizontal coverage plane $z = -H+h_{\text{cov}}$, where $1 \leq i \leq N_x$ and $1 \leq j \leq N_y$. Thus, we have
\begin{align}
	\mathbf{p}_{g}^{(i,j)} = \left( \left(i - \tfrac{1}{2}\right)\Delta x - \tfrac{L}{2}, \; \left(j - \tfrac{1}{2}\right)\Delta y - \tfrac{W}{2}, \; -H+h_{\text{cov}} \right),
\end{align}
with $\Delta x = \frac{L}{N_x}$ and $\Delta y = \frac{W}{N_y}$ being the width and height of each grid zone. Here, $h_{\text{cov}}$ denotes the vertical distance above the floor at which optical coverage is optimized, i.e., the height where the beam footprint is best aligned with the grid layout. This coverage design follows the 3D optical model and beam divergence analysis discussed in \cite{safi20253d}.

Each optical wireless transmission path is aligned to direct a photon beam from $\mathbf{p}_{\text{tx}}$ to the center of a corresponding grid. The direction vector of the beam toward grid $(i,j)$ is given by
\begin{align} \label{sq1}
	\mathbf{d}^{(i,j)} = \mathbf{p}_{g}^{(i,j)} - \mathbf{p}_{\text{tx}} = \mathbf{p}_{g}^{(i,j)} .
\end{align}
To ensure that the entire grid at $h_{\text{cov}}$ is covered by the optical beam, the divergence angle $\phi_{\text{div}}^{(i,j)}$ of each beam is adjusted based on its propagation vector $\mathbf{d}^{(i,j)}$. 
The divergence angle of each beam is tuned to ensure complete coverage at the receiver plane $z = h_{\text{cov}}$.
For further geometric considerations and coverage performance details, refer to \cite{safi20253d}.

Let $\mathbf{p}_{u} = [x_u, y_u, z_u]$ denote the true position of the user, and let $\hat{\mathbf{p}}_u = [\hat{x}_u, \hat{y}_u, \hat{z}_u]$ represent the estimated position available to the system. Here, we assume that the estimation is subject to random errors modeled as independent Gaussian noise in each coordinate. However, in the following subsections we investigate how alternative error models affect the performance of the synchronization system. Accordingly, we have
\begin{align}
	\hat{x}_u = x_u + n_x, \quad
	\hat{y}_u = y_u + n_y, \quad
	\hat{z}_u = z_u + n_z,
\end{align}
where $n_x, n_y, n_z \sim \mathcal{N}(0, \sigma_p^2)$ are zero-mean Gaussian random variables with identical variance $\sigma_p^2$. These represent the initial spatial uncertainty in user localization arising from sensing or tracking limitations.
Given $\hat{\mathbf{p}}_u$, the system activates the optical transmission path corresponding to the grid $(i^\star, j^\star)$ that minimizes the angular deviation between the beam direction $\mathbf{d}^{(i,j)}$ and the estimated user vector. Formally, the selected grid is given by
\begin{align}
	\label{selected_grid}
	(i^\star, j^\star) = \arg\min_{(i,j)} \; \theta_{\text{err}}^{(i,j)} = \arg\min_{(i,j)} \; \cos^{-1}\left( \frac{ \mathbf{d}^{(i,j)} \cdot \hat{\mathbf{d}}_u }{ \|\mathbf{d}^{(i,j)}\| \|\hat{\mathbf{d}}_u\| } \right),
\end{align}
where $\hat{\mathbf{d}}_u = \hat{\mathbf{p}}_u - \mathbf{p}_{\text{tx}}$ is the direction vector from the transmitter to the estimated user location. This selection ensures that the beam is directed toward the grid whose nominal axis most closely aligns with the estimated user position in 3D space.

Moreover, to simplify the analysis of photon propagation, we transform all global positions from the original $(x, y, z)$ coordinate system to a local coordinate frame $(x', y', z')$ aligned with the beam direction $\mathbf{d}^{(i,j)}$. In this frame, the $z'$-axis is parallel to $\mathbf{d}^{(i,j)}$, and the $x'$-$y'$ plane is perpendicular to it.
	The user position is then transformed from the global coordinate system to the beam-aligned local frame using
	\begin{align}
		\mathbf{p}_u' = \mathbf{R}^{(i,j)} \left( \mathbf{p}_u - \mathbf{p}_{\text{tx}} \right),
	\end{align}
	where $\mathbf{p}_u' = [x_u', y_u', z_u']^\top$ represents the user coordinates in the local $(x', y', z')$ frame, where $[\cdot]^\top$ denotes the transpose operator. The construction of the rotation matrix $\mathbf{R}^{(i,j)}$ is provided in Appendix~\ref{AppA}.

\subsection{Transmitter Architecture and Photon Pair Emission}
The central transmitter unit employs a SPDC source to generate polarization
encoded entangled photon pairs. Each generated pair consists of two photons: one designated as the \emph{reference photon} and the other as the \emph{user photon}. The reference photon is routed to a time-stamping detection module co-located at the transmitter, while the user photon is directed toward a spatially-resolved optical output based on the user's estimated position, as shown in Fig.~\ref{sm1}.

A SPDC source is widely adopted for quantum communication and synchronization due to its scalability and compatibility with free-space optics \cite{anwar2021entangled}. The system is driven by a periodic train of pump pulses, each with temporal width $\tau_p$. We model the photon-pair generation time $t_g$ as a Gaussian random variable centered at the pump pulse peak $t_p$, with standard deviation $\sigma_t$, as follows \cite{lasota2020optimal}
\begin{align} \label{eq:temporal-jitter}
	t_g \sim \mathcal{N}(t_p, \sigma_t^2).
\end{align}
This temporal uncertainty arises from the finite spectral bandwidth of the pump and the phase-matching conditions in the nonlinear crystal \cite{keller1997theory}. The jitter $\sigma_t$ represents an intrinsic timing error that limits the accuracy of photon-level synchronization. Typical values of $\sigma_t$ range from tens to hundreds of femtoseconds, depending on the pump-cavity configuration and spectral filtering \cite{allgaier2017fast}.
To ensure slot-level temporal isolation and mitigate timing ambiguity, each quantum slot is defined to be longer than the pump pulse duration, i.e., $T_{qs} > \tau_p$. Specifically, the slot duration is modeled as
\begin{align}
	T_{qs} = t_{\text{guard}}^{(1)} + \tau_p + t_{\text{guard}}^{(2)},
\end{align}
where $t_{\text{guard}}^{(1)}$ and $t_{\text{guard}}^{(2)}$ denote pre- and post-pulse guard intervals, respectively. These intervals are selected to absorb both the temporal jitter $\sigma_t$ of the SPDC source and the detection-time uncertainty $\sigma_d$ of the reference detector (e.g., SPAD).

Given the spontaneous nature of SPDC, the number of photon pairs generated within each quantum slot is random. Specifically, the emission statistics follow a Poisson distribution \cite{kim2022photon}
\begin{align} \label{eq:poisson-pair}
	\mathbb{P}(N_t = k) = \frac{\mu_t^k}{k!} e^{-\mu_t}, \quad \text{where } \mu_t < 1.
\end{align}
Here, $\mu_t$ denotes the mean number of photon pairs per slot.

Furthermore, each photon in the pair encodes a quantum bit (qubit) $|\psi\rangle$ in the polarization basis, defined as
\begin{align}
	|\psi\rangle = \alpha |0\rangle + \beta |1\rangle, \quad \text{with } |\alpha|^2 + |\beta|^2 = 1,
\end{align}
where $|0\rangle$ and $|1\rangle$ represent the horizontal and vertical polarization states, respectively. The entangled state of the pair can be written as:
\begin{align} \label{eq:epr1}
	|\Psi^{-}\rangle = \frac{1}{\sqrt{2}} \left( |0\rangle_r |1\rangle_u - |1\rangle_r |0\rangle_u \right)
\end{align}
where subscripts $r$ and $u$ denote the reference and user photons, respectively.

To direct the user photon toward the appropriate receiver location, the transmitter utilizes a dynamic beam selection mechanism based on estimated user coordinates $\hat{\mathbf{p}}_u$. As described in the previous subsection, the system activates the optical beam corresponding to the selected grid, minimizing the angular deviation between the estimated user direction and the grid axis (see \eqref{selected_grid}).

The switching mechanism at the transmitter steers the user photon to the beam path $\mathbf{d}^{(i^\star, j^\star)}$ via a high-speed optical switch array. Let $\mathcal{S}_{ij}$ denote the switch corresponding to path $(i,j)$. At each quantum slot, only one $\mathcal{S}_{ij}$ is activated. Simultaneously, the reference photon is routed internally to a time-reference block, where it is detected by a high-resolution single-photon detector (e.g., SPAD). The detection time $t_r$ is recorded and serves as a synchronization anchor for the corresponding user photon, allowing for precise time-of-arrival (ToA) estimation upon reception.

\subsection{Receiver Architecture and Photon Capture}

The user-side receiver, shown on the left-hand side of Fig.~\ref{sm1}, is designed to detect entangled photons transmitted via spatially directed beams from the central transmitter. The core component of the receiver is a circular aperture that captures incident photons based on their spatial alignment with the selected transmission beam. Each photon emitted toward the user is received with a probability $P_{\text{recept}}^{(i,j)}$, which depends on the alignment between the beam path $(i,j)$ and the user's actual position. The analytical model for $P_{\text{recept}}^{(i,j)}$ is developed in the following section.

After entering the aperture, the photon propagates through a passive optical path that directs it toward a detection module comprising an array of SPADs. These detectors are polarization-resolving and are used to determine the quantum state of the incoming photon based on its polarization encoding \cite{chandrasekharan2025resolving}. Each SPAD channel corresponds to a measurement basis (e.g., horizontal or vertical), allowing the receiver to recover the transmitted qubit from the spatially separated entangled photon pair.

Beyond merely detecting the presence of a photon, SPADs employed in this system are capable of precise time-tagging. Upon photon impact, an avalanche current is triggered within the diode, and the timestamp $t_d$ corresponding to the avalanche peak (or threshold crossing) is recorded. 
In practice, SPADs exhibit inherent temporal jitter due to both carrier avalanche buildup and electronic readout delays. This timing uncertainty is typically modeled as a Gaussian random variable $\sigma_d$ \cite{sun2019simple}, often referred to as the detector timing jitter
\begin{align} \label{eq:spad-jitter}
	t_d = t_a + n_d, \quad n_d \sim \mathcal{N}(0, \sigma_d^2),
\end{align}
where $t_a$ is the actual photon arrival time, and $t_d$ is the recorded detection time. The value of the standard deviation parameter, $\sigma_d$, depends on the SPAD's material and circuit design, and typically ranges from $20$~ps to $100$~ps in modern systems \cite{korzh2020demonstration,liu20241}. \textcolor{black}{The effective number of usable photon pairs available for synchronization is therefore jointly determined by the reception probability and the timestamp quality, which motivates the coupled spatial and temporal analysis developed in the following sections.}



\section{Spatial Modeling of Photon Reception via Aperture}
\label{modeling-sec}
The photon beam emitted by the transmitter is  directed toward the center of the selected grid $(i,j)$, following the predefined transmission vector $\mathbf{d}^{(i,j)}$ in \eqref{sq1}. 
Although the photon nominally propagates along $\mathbf{d}^{(i,j)}$, quantum diffraction causes its presence to be distributed laterally in the $x'$-$y'$ plane, with the probability density decaying radially from the beam axis \cite{leonhardt1997measuring}. This spatial uncertainty is captured by modeling the photon's transverse wavefunction as a Gaussian mode \cite{saleh2007fundamentals}. Specifically, the spatial probability density in the orthogonal plane at axial distance $z_u'$ is given by
\begin{align} \label{eq:pspd}
	P_{\text{spatial}}^{(i,j)}(x', y') = \frac{2}{\pi w^2(z_u')} \exp\left( -\frac{2(x'^2 + y'^2)}{w^2(z_u')} \right),
\end{align}
where $w(z_u')$ denotes the beam width at distance $z_u'$ along the propagation axis and is expressed as
\begin{align} \label{eq:bwid}
	w(z_u') = w_0 \sqrt{1 + \left( \frac{z_u' \lambda}{\pi w_0^2} \right)^2 },
\end{align}
with $w_0$ being the beam waist (minimum spot size at the focus), and $\lambda$ the photon wavelength.

The receiver aperture, centered at the user's position $\mathbf{p}_u' = [x_u', y_u', z_u']^\top$ in the beam-aligned coordinate frame, is assumed to lie in the plane orthogonal to the beam axis, i.e., the $x'$-$y'$ plane. The aperture captures photons based on their spatial distribution relative to the beam center.
Given the radial symmetry of the beam in the $x'$-$y'$ plane, the photon detection probability depends on the radial displacement of the user from the beam center. This displacement is quantified by the lateral components $x_u'$ and $y_u'$ of the user's position in the local frame.

Let $\rho_u' = \sqrt{(x_u')^2 + (y_u')^2}$ denote the radial offset of the receiver center from the beam axis. The photon detection probability is obtained by integrating the spatial density function \eqref{eq:pspd} over the receiver aperture, which is modeled as a circular disk of radius $r_a$ centered at $(x_u', y_u')$ in the $x'$-$y'$ plane. The detection probability is thus given by
\begin{align} \label{eq:pdet}
	P_{\text{recept}}^{(i,j)} = \iint_{\mathcal{A}_u} P_{\text{spatial}}^{(i,j)}(x', y') \, dx' \, dy',
\end{align}
where $\mathcal{A}_u$ denotes the circular region defined by
\begin{align} \label{eq:adsk}
	\mathcal{A}_u = \left\{ (x', y') \in \mathbb{R}^2 : (x' - x_u')^2 + (y' - y_u')^2 \leq r_a^2 \right\}.
\end{align}
When the beam width at the receiver plane is much larger than the aperture size, i.e., $w(z_u') > 4r_a$, the beam's intensity doesn't change much across the tiny area of the aperture \cite{safi2020analytical}. Thus, the spatial variation of the probability density across the receiver area becomes negligible. In this regime, the integral in \eqref{eq:pdet} can be approximated as 
\begin{align} \label{eq:pdap}
	P_{\text{recept}}^{(i,j)} \approx P_{\text{spatial}}^{(i,j)}(x_u', y_u') \cdot \pi r_a^2,
\end{align}
where $P_{\text{spatial}}^{(i,j)}(x_u', y_u')$ is evaluated using \eqref{eq:pspd}.

\textcolor{black}{It is also worth noting that, although independent Gaussian perturbations for spatial uncertainty are used as the baseline model for analytical tractability, the framework is not limited to Gaussian localization errors. In later sections, we also examine heavy tailed, biased, and spatially correlated position errors with matched root mean square deviation to assess robustness and to isolate the influence of spatial uncertainty on photon reception statistics. }

\section{Synchronization Model}
\label{synchro-sec}
In order to perform entanglement-based synchronization between the transmitter and the user receiver, we first define a discrete time frame consisting of a sequence of $N_s$ quantum time slots. Each slot is intended to carry a potential photon-pair emission by the source, with time-stamping performed by the reference unit located at the transmitter side. 

\subsection{Slot-Based Reference Sequence Construction}
Let $\mathcal{S} = \{1, 2, \dots, N_s\}$ denote the set of all quantum slots, and for each slot $k \in \mathcal{S}$, let $N_t[k]$ denote the number of photon pairs generated. As described earlier in \eqref{eq:poisson-pair}, this number follows a Poisson distribution with mean $\mu_t < 1$.
For synchronization purposes, the reference unit maintains two sequences as follows:
\begin{itemize}
	\item A data detection sequence $\mathbf{d}_r = [d_r[1], d_r[2], \dots, d_r[N_s]]$
	\item A timestamp sequence $\mathbf{t}_r = [t_r[1], t_r[2], \dots, t_r[N_s]]$
\end{itemize}
where $d_r[k] \in \{0,1,\texttt{fail}\}$ indicates the detection status of a valid reference photon in slot $k$, and $t_r[k]$ is the recorded timestamp when $d_r[k]=1$.
A quantum time slot $k$ is marked as \texttt{fail} under the two following conditions:
\begin{itemize}
	\item No photon pair was generated ($N_t[k] = 0$)
	\item More than one photon pair was generated ($N_t[k] \geq 2$)
\end{itemize}
Accordingly, only those slots for which $N_t[k] = 1$ are considered valid and eligible for synchronization. The rest are marked as unreliable. Therefore, we have
\begin{align}
	d_r[k] = 
	\begin{cases}
		1, & \text{if } N_t[k] = 1 \text{ and photon detected}, \\
		0, & \text{if } N_t[k] = 1 \text{ and no photon detected}, \\
		\texttt{fail}, & \text{if } N_t[k] = 0 \text{ or } N_t[k] \geq 2.
	\end{cases} \label{eq:dr-status}
\end{align}

Due to physical limitations of the SPAD detector at the reference unit, the timestamp $t_r[k]$ is not the exact emission time but includes additive Gaussian noise. Specifically, for each slot $k$ with $d_r[k] = 1$, the recorded time is
\begin{align}
	t_r[k] = t_g[k] + n_r[k], \quad n_r[k] \sim \mathcal{N}(0, \sigma_d^2) \label{eq:ref-timestamp}
\end{align}
where $t_g[k]$ is the true photon generation time for slot $k$, and $n_r[k]$ models the detection-time jitter at the reference SPAD.
It is important to note that the generation time $t_g[k]$ itself is also a random variable due to intrinsic temporal uncertainty of the SPDC process, modeled in \eqref{eq:temporal-jitter}. Thus, we have
\begin{align}
	t_g[k] \sim \mathcal{N}(t_p[k], \sigma_t^2) \label{eq:gen-time}
\end{align}
where $t_p[k]$ is the center of the pump pulse for slot $k$.
Consequently, the effective timestamp $t_r[k]$ observed at the reference is subject to a compound noise
\begin{align}
	t_r[k] \sim \mathcal{N}(t_p[k], \sigma_t^2 + \sigma_d^2) \label{eq:ref-noise-total}
\end{align}
which combines the SPDC generation jitter and the SPAD detection jitter. In contrast, the detection noise at the user's SPAD is assumed independent and modeled separately.

\subsection{User-Side Detection Sequence}
In each quantum time slot $k$, the number of signal photons $N_s[k]$ arriving at the aperture of the user receiver follows a Poisson distribution. Its probability mass function is given by
\begin{align}
	\mathbb{P}(N_s[k] = n) = \frac{(\mu_s^{(i,j)})^n}{n!} e^{-\mu_s^{(i,j)}}, \quad n = 0, 1, 2, \dots \label{eq:poisson-sig}
\end{align}
where the mean value $\mu_s^{(i,j)}$ depends on three factors: the mean photon-pair generation rate $\mu_t$ at the transmitter (see \eqref{eq:poisson-pair}), the photon reception probability $P_{\text{recept}}^{(i,j)}$ determined by spatial alignment (see \eqref{eq:pdap}), and the photon detection efficiency $\eta_d$ of the user's SPAD detector. Specifically,
\begin{align}
	\mu_s^{(i,j)} = \mu_t \cdot P_{\text{recept}}^{(i,j)} \cdot \eta_d. \label{eq:mu-sig}
\end{align}

In addition to signal photons, the receiver also detects background photons that originate from ambient light sources, thermal noise, or detector dark counts. These events are independent of the entangled source and are modeled as a separate Poisson process with mean arrival rate $\mu_b$. Because background photons have random polarization states, each such photon has a probability of $1/2$ to appear in either polarization-sensitive SPAD channel.
Therefore, the total number of photons impinging on the receiver in slot $k$ is the sum of signal and background counts
\begin{align}
	N_u[k] = N_s[k] + N_b[k], \label{eq:nuser}
\end{align}
where $N_s[k] \sim \text{Poisson}(\mu_s^{(i,j)})$ and $N_b[k] \sim \text{Poisson}(\mu_b)$ are independent.

At the user receiver, entangled photons are captured and registered over the same quantum slot structure $\mathcal{S} = \{1, 2, \dots, N_s\}$ as defined at the reference side. The receiver constructs a binary detection sequence $\mathbf{d}_u = [d_u[1], d_u[2], \dots, d_u[N_s]]$, where each element $d_u[k] \in \{0, 1, \texttt{fail}\}$ reflects the outcome of photon detection and polarization measurement during slot $k$.
Let us define the detection bit $d_u[k]$ for slot $k$ as follows:
\begin{itemize}
	\item $d_u[k] = 1$ if exactly one signal photon is detected and its polarization is correctly resolved.
	\item $d_u[k] = 0$ if a signal photon is detected but the measured polarization is incorrect (due to overlap with background).
	\item $d_u[k] = \texttt{fail}$ in all other cases: either no photon was detected or multiple photons caused measurement ambiguity.
\end{itemize}
A slot is most likely to result in a \texttt{fail} status due to the absence of any detected photons (i.e., $N_u[k] = 0$). However, even if a signal photon is present, a \texttt{fail} can also occur if:
\begin{enumerate}
	\item A background photon arrives simultaneously with a signal photon ($N_s[k]=1$, $N_b[k] \geq 1$)
	\item The background photon's polarization differs from that of the signal, leading to a contradiction in polarization detection.
\end{enumerate}

In quantum slots where $d_u[k] \in \{0, 1\}$ (i.e., photon detection is successful and not marked as \texttt{fail}), the user receiver records the photon arrival time $t_u[k]$. This timestamp captures three main factors:
\begin{itemize}
	\item The photon generation time $t_g[k]$ at the transmitter (see \eqref{eq:gen-time})
	\item A cumulative synchronization deviation $\tau_{\text{sync}}$, which includes both the physical propagation delay and any initial offset due to system synchronization mismatch
	\item The SPAD detector’s timing jitter $n_u[k] \sim \mathcal{N}(0, \sigma_d^2)$, modeled as Gaussian noise
\end{itemize}
Accordingly, for each valid detection slot $k$, the timestamp recorded at the user is given by
\begin{align}
	t_u[k] = t_g[k] + \tau_{\text{sync}} + n_u[k], \quad n_u[k] \sim \mathcal{N}(0, \sigma_d^2). \label{eq:tu-model}
\end{align}

Here, $t_g[k]$ is itself a Gaussian-distributed variable centered around the pump pulse time $t_p[k]$ (as described in \eqref{eq:gen-time}). Thus, the total timing uncertainty at the receiver combines the intrinsic generation jitter $\sigma_t$ and the detector-side jitter $\sigma_d$, resulting in the following distribution
\begin{align}
	t_u[k] \sim \mathcal{N}(t_p[k] + \tau_{\text{sync}}, \sigma_t^2 + \sigma_d^2). \label{eq:tu-distribution}
\end{align}
Finally, the complete user-side timestamp sequence is represented as
\begin{align}
	\mathbf{t}_u = [t_u[1], t_u[2], \dots, t_u[N_s]],
\end{align}
where each element $t_u[k]$ is defined only for slots where $d_u[k] \in \{0, 1\}$, and marked as \texttt{fail} otherwise.

\begin{algorithm}[t]
	\caption{Quantum Synchronization via Sparse Bit Matching}
	\label{alg:sync}
	\begin{algorithmic}[1]
		\State \textbf{Input:} User detection vector $\mathbf{d}_u$, user timestamps $\mathbf{t}_u$, reference detection vector $\mathbf{d}_r$, reference timestamps $\mathbf{t}_r$
		\State \textbf{Output:} Estimated synchronization offset $\hat{\tau}_{\text{sync}}$
		\vspace{0.5em}
		\State Extract index set $\mathbf{n}_u = \{i_1, \dots, i_x\}$ such that $d_u[i_j] \in \{0,1\}$
		\State Extract bit values $\mathbf{v}_u = [d_u[i_1], \dots, d_u[i_x]]$
		\For{each shift $\ell$ in candidate set $\mathcal{L}$}
		\State $S[\ell] \gets 0$
		\For{$j = 1$ to $x$}
		\If{$d_r[i_j - \ell] \in \{0,1\}$ and $d_r[i_j - \ell] = d_u[i_j]$}
		\State $S[\ell] \gets S[\ell] + 1$
		\EndIf
		\EndFor
		\EndFor
		\State $\ell^\star \gets \arg\max_{\ell \in \mathcal{L}} S[\ell]$
		\State Construct matching index set $\mathcal{J}_{\text{match}} = \{ j : d_r[i_j - \ell^\star] = d_u[i_j] \}$
		\State $\hat{\tau}_{\text{sync}} \gets \frac{1}{|\mathcal{J}_{\text{match}}|} \sum_{j \in \mathcal{J}_{\text{match}}} (t_u[i_j] - t_r[i_j - \ell^\star])$
		\State \textbf{return} $\hat{\tau}_{\text{sync}}$
	\end{algorithmic}
\end{algorithm}

\subsection{Synchronization Recovery at the User}
To establish a shared temporal reference between the transmitter and the user receiver, the system utilizes a classical auxiliary synchronization channel in addition to the quantum photon link. This classical link, implemented using technologies such as  vertical-cavity surface-emitting laser (VCSEL)-based narrow optical beams, is made feasible by the structured beam-steering setup already present at the transmitter \cite{zeng2021vcsel}. Through this channel, the transmitter broadcasts two sequences aligned with the quantum time slots, i.e., the binary detection sequence $\mathbf{d}_r = [d_r[1], \dots, d_r[N_s]]$ and the associated timestamp sequence $\mathbf{t}_r = [t_r[1], \dots, t_r[N_s]]$, which represent detection outcomes and timing information of reference photons, respectively.

At the user side, upon receiving photons through the quantum aperture, the receiver builds its own detection sequence $\mathbf{d}_u = [d_u[1], \dots, d_u[N_s]]$ and timestamp vector $\mathbf{t}_u = [t_u[1], \dots, t_u[N_s]]$, as previously described. However, due to the unknown synchronization offset $\tau_{\text{sync}}$ between local and transmitter clocks, the timestamps in $\mathbf{t}_u$ are shifted relative to those in $\mathbf{t}_r$. This alignment is further complicated by the fact that a significant portion of transmitted photons result in \texttt{fail} states at the receiver, primarily due to the combined effects of source-side timing jitter, SPAD detector noise, and interference from background photons. As a result, only a small subset of quantum slots provides reliable and temporally matched detection events, making accurate synchronization estimation more challenging under practical conditions.

Due to the random nature of photon detection, the binary sequence $\mathbf{d}_u = [d_u[1], \dots, d_u[N_s]]$ constructed at the user side includes a large number of \texttt{fail} entries, corresponding to undetected or ambiguous photon events. In contrast, the reference sequence $\mathbf{d}_r = [d_r[1], \dots, d_r[N_s]]$ generated at the transmitter is structurally aligned with the quantum slot framework, although it may also contain \texttt{fail} entries due to multi-photon emissions or null generation events. Nevertheless, the fail rate in $\mathbf{d}_r$ is typically much lower than that observed in the user-side sequence $\mathbf{d}_u$. The primary challenge is that the valid bits in $\mathbf{d}_u$ represent only a sparse and potentially misaligned sample of the original transmission pattern. Additionally, some valid detections at the user may be corrupted due to background-induced bit errors.

Considering the above-mentioned challenges, and to enable robust synchronization, we extract two auxiliary vectors from $\mathbf{d}_u$ as follows
\begin{itemize}
	\item A valid data vector $\mathbf{v}_u = [d_u[i_1], d_u[i_2], \dots, d_u[i_x]]$, consisting of all elements where $d_u[i_j] \in \{0,1\}$,
	\item A corresponding index vector $\mathbf{n}_u = [i_1, i_2, \dots, i_x]$, recording the slot positions of these valid entries.
\end{itemize}
We then perform pattern matching between the observed vector $\mathbf{v}_u$ and shifted versions of the transmitter-side reference vector $\mathbf{d}_r$, evaluating how well $\mathbf{v}_u$ aligns with subsequences of $\mathbf{d}_r$. For each candidate integer shift $\ell \in \mathcal{L}$, we compute a matching score as
\begin{align}
	S[\ell] = \sum_{j=1}^{x} \delta\left( d_u[i_j], d_r[i_j - \ell] \right), \label{eq:matching-score}
\end{align}
where $\delta(a,b) = 1$ if $a = b$, and $0$ otherwise. The optimal alignment offset is then estimated as
\begin{align}
	\ell^\star = \arg\max_{\ell \in \mathcal{L}} S[\ell], \label{eq:best-shift}
\end{align}
which identifies the temporal shift that yields the highest agreement between the user and reference detection sequences, ignoring \texttt{fail} entries.

Once the optimal shift $\ell^\star$ is identified, the user constructs a refined index set $\mathcal{J}_{\text{match}}$ by filtering those valid entries for which both transmitter and receiver have defined and matching binary values
\begin{align}
	&\mathcal{J}_{\text{match}} = \\
	&\left\{ j \in \{1,\dots,x\} \;:\; d_r[i_j - \ell^\star] \in \{0,1\},\; d_r[i_j - \ell^\star] = d_u[i_j] \right\}. \nonumber
\end{align}
This filtered subset ensures that the timestamp estimation only uses reliably matched positions, where both the quantum event occurred and the polarization was correctly inferred at the receiver.
The final synchronization offset is then estimated by computing the average temporal difference over the selected indices, i.e., 
\begin{align}
	\hat{\tau}_{\text{sync}} = \frac{1}{|\mathcal{J}_{\text{match}}|} \sum_{j \in \mathcal{J}_{\text{match}}} \left( t_u[i_j] - t_r[i_j - \ell^\star] \right). \label{eq:sync-est-final}
\end{align}
This two-stage matching and refinement process enables robust synchronization even in the presence of asymmetric detection failures, background-induced bit flips, and missing reference values. It ensures that only high-confidence matchings contribute to the final clock alignment.

Based on the statistical detection and timing models defined in \eqref{eq:tu-model}–\eqref{eq:tu-distribution} and the matching score formulation in \eqref{eq:matching-score}–\eqref{eq:best-shift}, the proposed synchronization strategy is summarized in Algorithm~\ref{alg:sync}. This procedure leverages a two-stage approach. First, it estimates the most likely alignment shift, $\ell^\star$, between the sparse user-side detection vector $\mathbf{d}_u$ and the reference pattern $\mathbf{d}_r$ using discrete correlation. Then, it constructs a refined set of matching slots $\mathcal{J}_{\text{match}}$ that excludes ambiguous or non-informative events. The synchronization offset $\hat{\tau}_{\text{sync}}$ is subsequently computed via averaging over timestamp differences as described in \eqref{eq:sync-est-final}. This method allows reliable clock recovery even under heavy detection sparsity and measurement noise, and forms the basis for slot-level temporal alignment required for entanglement correlation and key distribution in quantum communication scenarios.


\section{Fundamental Limits: Spatio-Temporal Cramér-Rao Bound}
\label{sec:crlb}

\textcolor{black}{To establish the fundamental performance limits of our synchronization framework and demonstrate the novel spatio-temporal coupling introduced by the grid architecture, we derive the CRLB for the synchronization offset estimate $\hat{\tau}_{\text{sync}}$. This analysis reveals how spatial grid parameters directly influence the achievable timing precision, a distinctive feature of our quantum wireless system that has no classical counterpart.}

\textcolor{black}{\subsubsection{Likelihood Function for Timestamp Differences}
After successful bit-pattern matching (Algorithm~\ref{alg:sync}), we obtain a set $\mathcal{J}_{\text{match}}$ of reliably matched slots with cardinality $M = |\mathcal{J}_{\text{match}}|$. For each matched slot $j$, we observe the timestamp difference as
\begin{equation}
	\Delta_j = t_u[i_j] - t_r[i_j - \ell^*],
	\label{delta_CRLB}
\end{equation}
where $t_r[k]$ and $t_u[k]$ are given by~\eqref{eq:ref-timestamp} and~\eqref{eq:tu-model}, respectively. The reference timestamp is for a successfully detected photon pair. Accordingly, one can rewrite \eqref{delta_CRLB} as
\begin{equation}
	\Delta_j = \tau_{\text{sync}} + (n_u[i_j] - n_r[i_j - \ell^*]).
	\label{delta_CRLB2}
\end{equation}
Given the independence of $n_u$ and $n_r$, the difference in \eqref{delta_CRLB2} follows
\begin{equation}
	\Delta_j \sim \mathcal{N}(\tau_{\text{sync}}, 2\sigma_d^2).
\end{equation}
The likelihood function for $M$ independent observations can thus be written as
\begin{equation}
	f(\{\Delta_j\}_{j=1}^M | \tau_{\text{sync}}) = \prod_{j=1}^M \frac{1}{\sqrt{4\pi\sigma_d^2}} \exp\left(-\frac{(\Delta_j - \tau_{\text{sync}})^2}{4\sigma_d^2}\right).
\end{equation}}
\textcolor{black}{\subsubsection{Fisher Information and CRLB}
Let’s write the log-likelihood function with respect to the synchronization delay parameter $\tau_{\text{sync}}$ as
\begin{equation}
	\mathcal{L}(\tau_{\text{sync}}) = -\frac{M}{2}\ln(4\pi\sigma_d^2) - \sum_{j=1}^M \frac{(\Delta_j - \tau_{\text{sync}})^2}{4\sigma_d^2}.
\end{equation}
The Fisher Information for $\tau_{\text{sync}}$ is then obtained as 
\begin{equation}
	I(\tau_{\text{sync}}) = -\mathbb{E}\left[\frac{\partial^2 \mathcal{L}}{\partial \tau_{\text{sync}}^2}\right] = \sum_{j=1}^M \frac{1}{2\sigma_d^2} = \frac{M}{2\sigma_d^2}.
\end{equation}
Thus, the CRLB for any unbiased estimator $\hat{\tau}_{\text{sync}}$ is
\begin{equation}
	\label{eq:crlb_base}
	{\text{Var}(\hat{\tau}_{\text{sync}}) \geq \frac{1}{I(\tau_{\text{sync}})} = \frac{2\sigma_d^2}{M}.}
\end{equation}
This bound holds for any unbiased estimator conditioned on $M$ successfully matched photon pairs and shows that the achievable timing precision scales inversely with the number of reliable detections. }

\textcolor{black}{\subsubsection{Spatial Dependence of Matched Pair Count}
The number of matched pairs $M$ is a random variable whose expectation depends critically on spatial parameters. More precisely, for a synchronization sequence comprising 
$N_{s} = T_{\text{seq}} / T_{\text{qs}}$ slots, a valid single‑pair emission occurs with probability 
$\mu_{t} e^{-\mu_{t}}$ under Poisson statistics. Conditioning on a single‑pair event, a successful match further requires photon reception, detection, and the absence of background‑induced ambiguity. Hence, the expected number is
\begin{equation}
	\label{expected_m}
	\mathbb{E}[M] = N_s \cdot \mu_{t} e^{-\mu_{t}} \cdot \overline{P}_{\text{match}}(N_g, \sigma_p),
\end{equation}
where  $\overline{P}_{\text{match}}(N_g, \sigma_p)$ is the average probability that a generated photon pair results in a successful match.
This matching probability incorporates spatial alignment effects through
\begin{equation}
	\overline{P}_{\text{match}}(N_g, \sigma_p) = \eta_d (1 - p_{\text{back}}) \cdot \mathbb{E}[P_{\text{recept}}^{(i,j)}],
\end{equation}
where $\eta_d$ is the detector efficiency, $p_{\text{back}}$ accounts for background-induced failures, i.e., the probability that background photons cause a slot to be invalid for synchronization. Also, $\mathbb{E}[P_{\text{recept}}^{(i,j)}]$ is the expected reception probability over user position uncertainty (using the spatial reception model developed in Section \ref{modeling-sec}).
}
\textcolor{black}{From~\eqref{eq:pdap}, the reception probability for a beam aligned with grid $(i,j)$ is
\begin{equation}
	P_{\text{recept}}^{(i,j)} \approx  \frac{2r_a^2}{ w^2(z_u')} \exp\left(-\frac{2[(x_u')^2 + (y_u')^2]}{w^2(z_u')}\right),
\end{equation}
where $(x_u', y_u')$ represents the user's offset from the beam center in local coordinates. With Gaussian position uncertainty $\sigma_p^2$ in each coordinate, the expected reception probability becomes
\begin{equation}
	\label{eq:expected_precept}
	\mathbb{E}[P_{\text{recept}}^{(i,j)}] = \frac{2r_a^2}{w^2(z_u') + 4\sigma_p^2}.
\end{equation}
Finally, combining~\eqref{eq:crlb_base},~\eqref{eq:expected_precept}, and \eqref{expected_m}, we obtain the CRLB that yields the following information-theoretic bound 
\begin{equation}
	\label{beam_prop_ex}
	\mathrm{Var}(\hat{\tau}_{\mathrm{sync}})
	\geq
	\frac{\sigma_d^2}{N_s \mu_{t} e^{-\mu_{t}} \eta_d \left(1-p_{\mathrm{back}}\right)}
	\cdot
	\frac{w^2(z_u') + 4\sigma_p^2}{r_a^2}.
\end{equation}}
\textcolor{black}{This bound reveals a direct coupling between spatial beam design and temporal synchronization accuracy. In particular, reducing the beam width $w(z_u')$ through finer grid resolution improves timing precision only until $w^{2}(z_u')$ becomes comparable to the position‑uncertainty term $4\sigma_{p}^{2}$. Beyond this point, further grid refinement yields diminishing returns, as synchronization performance becomes dominated by localization uncertainty rather than beam focusing.}
	
\textcolor{black}{
This behavior highlights a fundamental distinction between classical and quantum grid‑based systems. While classical GoB architectures primarily trade spatial resolution for received optical power, in quantum synchronization systems the spatial architecture explicitly enters the information bound governing timing accuracy. To the best of our knowledge, this spatio‑temporal coupling has not been previously characterized in the context of indoor entanglement‑assisted synchronization.}

{\section{Channel Impairments and Timing Uncertainty Model}
\label{sec:channel_impairments}

\textcolor{black}{In the previous sections, idealized channel conditions were assumed in order to isolate the fundamental operation of the synchronization framework. In this section, we introduce the channel impairments considered in the proposed framework and formalize their impact on the timing observations used for synchronization. The purpose of this section is to define the physical mechanisms that degrade timing accuracy in indoor optical wireless environments, independently of the specific synchronization algorithm.}

\textcolor{black}{\subsection{Multipath Propagation and Non-Line-of-Sight Components}}

\textcolor{black}{In indoor optical wireless systems, photons can reach the receiver through both a direct line-of-sight (LOS) path and multiple reflected paths arising from walls, ceilings, and surrounding objects. While the LOS component carries the correct timing reference, reflected components arrive with excess delay and introduce temporal dispersion.}
\textcolor{black}{To model this effect, the optical channel impulse response is represented using a double-exponential form commonly adopted for indoor optical wireless channels \cite{carruthers2002iterative,al2018optical,akhoundi2015cellular}:
\begin{align}
	h_{\text{total}}(t) =
	\eta_{\text{LOS}} \delta(t)
	+ \sum_{k=1}^{N_{\text{ref}}} \eta_k
	\exp\!\left(-\frac{t - \Delta t_k}{\tau_{\text{rms}}}\right),
\end{align}
where $\eta_{\text{LOS}}$ denotes the weight of the LOS component, $\eta_k$ represents the contribution of the $k$-th reflected path, $\Delta t_k$ is its associated excess delay, and $\tau_{\text{rms}}$ is the root-mean-square (RMS) delay spread of the channel.}

\textcolor{black}{Here,  we assume that the LOS term dominates in typical indoor environments; however, reflected components introduce a non-negligible tail in the impulse response \cite{al2018optical}. We also assume the RMS delay spread is set to $\tau_{\text{rms}} = 300$~ps, consistent with reported indoor optical measurements \cite{ghassemlooy2013indoor}. This multipath induced dispersion introduces a systematic bias in the detected photon arrival times and represents a fundamental limitation on synchronization accuracy, which we examine in detail in the performance evaluation section.}

\textcolor{black}{\subsection{Non-Gaussian Timing Jitter and Diffusion Effects}}

\textcolor{black}{In addition to channel induced dispersion, the temporal resolution of the receiver is constrained by the intrinsic timing jitter of the SPAD. A Gaussian model was used in the previous section for analytical tractability. However, although Gaussian jitter models are commonly adopted for convenience, experimental studies have shown that SPAD timing responses exhibit pronounced long tailed behavior caused by carrier diffusion within the absorption and multiplication regions of the detector \cite{sun2019simple}.}

\textcolor{black}{To capture this effect and to assess the robustness of the proposed synchronization framework under realistic conditions, we also model the detector timing jitter using a Laplacian distribution and provide the performance evaluation. This choice is physically justified by the exponential nature of diffusion-induced delays and has been widely adopted in photon-timing experiments \cite{yu2024recent,hadfield2009single}. The corresponding probability density function is given by
\begin{align}
	f(x \mid \mu, b) = \frac{1}{2b} \exp\!\left(-\frac{|x-\mu|}{b}\right),
\end{align}
where $\mu$ denotes the nominal arrival time and the scale parameter $b$ is related to the standard deviation $\sigma$ via $b = \sigma / \sqrt{2}$.}

\textcolor{black}{The Laplacian model exhibits heavier tails than a Gaussian distribution and therefore provides a conservative representation of timing uncertainty. To have a fair comparison, in this work, the Laplacian parameters are selected such that the variance matches that of the Gaussian jitter model used in the baseline case in the previous section.}

\textcolor{black}{\subsection{Unified Timing Observation Model}}

\textcolor{black}{Taking into account photon generation uncertainty, multipath dispersion, and detector timing jitter, the timestamp recorded at the receiver (i.e., \eqref{eq:tu-model}) can be expressed as
\begin{align}
	t_u = t_g + \tau_{\mathrm{sync}} + \Delta t_{\mathrm{multipath}} + n_{\mathrm{jitter}},
\end{align}
where $t_g$ is the photon generation time, $\tau_{\mathrm{sync}}$ denotes the clock offset between the transmitter and receiver, $\Delta t_{\mathrm{multipath}}$ represents the excess delay caused by reflected paths, and $n_{\mathrm{jitter}}$ follows the Laplacian distribution defined above. This formulation explicitly separates deterministic synchronization offsets from stochastic channel and detector effects. Importantly, it highlights that multipath propagation introduces a bias term that does not vanish with increasing photon count, while detector jitter determines the convergence behavior of the synchronization estimator. These effects are examined quantitatively in the performance evaluation section, where their impact on synchronization accuracy and convergence behavior is assessed.}

\section{Simulations}
\label{simulations-sec}
To assess the synchronization performance under realistic quantum indoor environments, we conducted Monte-Carlo simulations based on the system model described in Sections \ref{sysmode-sec}–\ref{synchro-sec}. The room was configured as a rectangular indoor space with dimensions $L = 6\,\text{m}$, $W = 6\,\text{m}$, and $H = 3\,\text{m}$. The quantum optical coverage was set at height $h_{\text{cov}} = 1\,\text{m}$ above the floor, and the grid-based beam steering was evaluated for multiple spatial configurations with $N_g \in \{25, 100, 225\}$, corresponding to $5 \times 5$, $10 \times 10$, and $15 \times 15$ grid layouts  \cite{safi20253d}. Each grid region was targeted by a distinct optical beam, and photon propagation followed the diffraction and spatial uncertainty models described earlier. 

The synchronization sequence spanned a total duration $T_{\text{seq}}$ ranging from $1\,\mu\text{s}$ to $2\,\text{ms}$, divided into slots of duration $T_{qs} = 10\,\text{ns}$. This results in a total number of time slots $N_s = T_{\text{seq}} / T_{qs}$, which varied between $10^2$ and $2 \times 10^5$ depending on the experiment. Each slot could contain a single entangled photon pair with mean generation rate $\mu_t = 0.5$, while the SPAD detector timing jitter was modeled as Gaussian noise with standard deviation $\sigma_d = 200\,\text{ps}$. The SPDC source temporal jitter was modeled with standard deviation $\sigma_t = 50\,\text{ps}$ \cite{quan2016demonstration}. Background photon arrivals followed a Poisson distribution with mean $\mu_b = 5 \times 10^{-6}$ \cite{bronzi2015spad}. The quantum detection efficiency of the SPAD was set to $\eta_d = 0.6$, and the user receiver was modeled with a circular aperture of radius $r_a = 0.02\,\text{m}$. Unless otherwise stated, the simulations assume an LOS channel and use the Gaussian detector jitter model.
The simulation parameters are summarized in Table~\ref{tab:sim_params}, where variables marked with a range were evaluated parametrically, and exact values for each scenario are specified in the respective figure captions.

\begin{table}[H]
	\centering
	\caption{Simulation Parameters for Quantum Synchronization System}
	\label{tab:sim_params}
	\begin{tabular}{|l|l|l|}
		\hline
		\textbf{Parameter} & \textbf{Symbol} & \textbf{Value / Range} \\
		\hline
		Room length & $L$ & $6\,\text{m}$ \\
		Room width & $W$ & $6\,\text{m}$ \\
		Room height & $H$ & $3\,\text{m}$ \\
		Coverage height & $h_{\text{cov}}$ & $1\,\text{m}$ \\
		Number of grids & $N_g$ & $25$, $100$, $225$ \\
		Beam width per grid & $w_z^{(i,j)} $ & $= \frac{W}{N_x}$ \\
		Receiver aperture radius & $r_a$ & $0.02\,\text{m}$ \\
		Slot duration & $T_{qs}$ & $10\,\text{ns}$ \\
		Synchronization duration & $T_{\text{seq}}$ & $1$–$2000\,\mu\text{s}$ \\
		Number of time slots & $N_s$ & $10^2$–$2 \times 10^5$ \\
		Mean photon-pair rate per slot & $\mu_t$ & $0.5$ \\
		SPAD detection efficiency & $\eta_d$ & $0.6$ \\
		Mean background photons per slot & $\mu_b$ & $5 \times 10^{-6}$ \\
		SPDC temporal jitter & $\sigma_t$ & $50\,\text{ps}$ \\
		SPAD timing jitter & $\sigma_d$ & $200\,\text{ps}$ \\
		Initial position error std. dev. & $\sigma_p$ & $6\,\text{cm}$ \\
		Delay spred & $\tau_{\text{rms}}$ & $300\,\text{ps}$ \\
		\hline
	\end{tabular}
\end{table}





\subsection{Baseline Synchronization Performance under Ideal Conditions}

\begin{figure*}
	\centering
	\subfloat[] {\includegraphics[width=3.4 in]{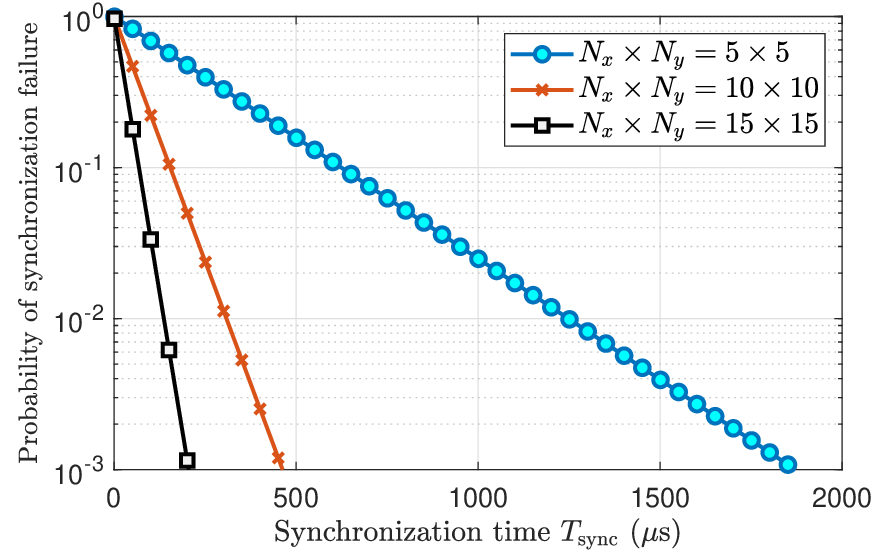}
		\label{gb1}
	}
	\hfill
	\subfloat[] {\includegraphics[width=3.4 in]{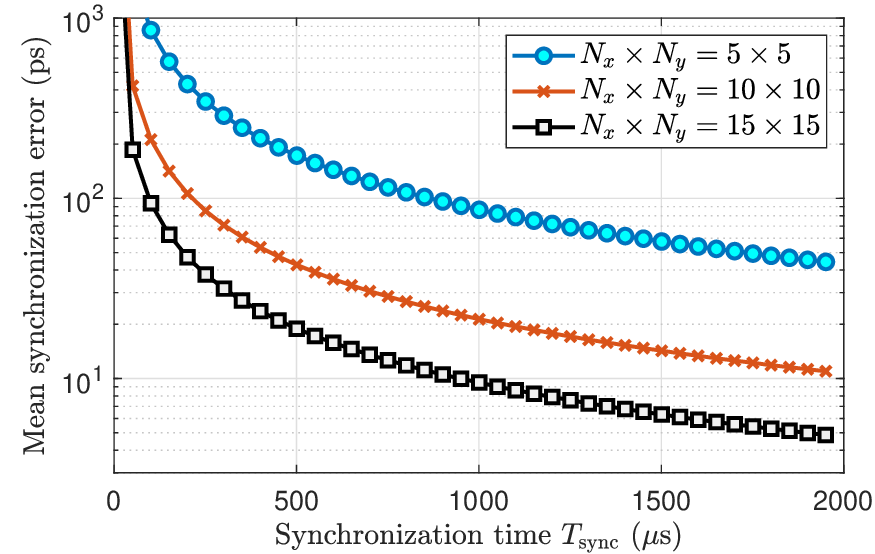}
		\label{gb2}
	}
	\caption{Effect of synchronization sequence duration $T_{\text{seq}}$ and grid configuration on synchronization performance. (a) Probability of synchronization failure and (b) Average synchronization error versus $T_{\text{seq}}$ under different spatial grid sizes. A failure occurs when fewer than two valid detection bits are registered within the allotted duration, making offset estimation infeasible.}
	
	\label{gb}
\end{figure*}

Fig.~\ref{gb} evaluates the effect of synchronization sequence duration $T_{\text{seq}}$ on the quantum synchronization performance under different spatial grid configurations. Each subfigure corresponds to a distinct performance metric, with subfigure (a) showing the probability of synchronization failure, and subfigure (b) reporting the average synchronization error between the user and transmitter time references.
As illustrated in Fig.~\ref{gb1}, the probability of synchronization failure, which is defined as the event where fewer than two valid detection bits are observed at the user within the allotted synchronization time, decreases monotonically with increasing $T_{\text{seq}}$. This phenomenon arises due to the inherent sparsity and stochastic nature of photon reception in wireless quantum systems. Given the probabilistic losses in entangled photon transmission and background interference, many quantum slots yield no valid detections. When the synchronization duration is insufficient, the receiver may fail to collect the minimum required number of valid qubits (two or more), leading to a failure in offset estimation.
Fig.~\ref{gb2} further illustrates that the mean synchronization error, which is defined as the average deviation between the estimated synchronization offset $\hat{\tau}_{\text{sync}}$ and the true offset $\tau_{\text{sync}}$, declines significantly as $T_{\text{seq}}$ increases. A longer sequence provides more detection opportunities, enhancing the likelihood of high-confidence bit matching and robust timestamp averaging.
Additionally, both metrics demonstrate that finer spatial grid configurations (e.g., $15 \times 15$) lead to better synchronization outcomes compared to coarser grids (e.g., $5 \times 5$). This improvement is due to enhanced beam alignment, which increases the photon reception probability $P_{\text{recept}}^{(i,j)}$ (as derived in \eqref{eq:pdap}). With narrower beam footprints, users are more likely to fall within high-intensity zones, enabling valid qubit detection even under shorter synchronization durations. For instance, after $T_{\text{seq}} = 1\,\text{ms}$, the average synchronization error is reduced from approximately $90\,\text{ps}$ in the $5 \times 5$ case to below $9\,\text{ps}$ in the $15 \times 15$ configuration.


Another critical parameter that significantly affects the performance of wireless quantum synchronization systems is the mean photon-pair generation rate per slot, denoted by $\mu_t$. Fig.~\ref{fml1} illustrates the average synchronization error as a function of $\mu_t$ over a wide range from $0.1$ to $4$, under fixed environmental and timing configurations. The simulation considers a $15 \times 15$ grid structure ($N_g = 225$) and a synchronization duration of $T_{\text{seq}} = 1\,\text{ms}$ per user. All other system parameters follow the nominal values specified in Table~\ref{tab:sim_params}.
As evident in the results, the value of $\mu_t$ plays a dual and non-trivial role in synchronization accuracy. When $\mu_t$ is too small, the probability that the SPDC source generates a valid single photon pair in a given quantum slot decreases sharply, leading to high rates of slot failure and consequently degraded synchronization performance due to insufficient valid timestamps. Conversely, if $\mu_t$ becomes too large, the probability of generating two or more photon pairs in the same slot increases, which, according to the protocol defined in \eqref{eq:dr-status}, renders the slot invalid for synchronization purposes. Such multi-pair emissions introduce ambiguity and break the one-to-one correspondence between reference and user photons.
Therefore, there exists an optimal range for $\mu_t$ that balances the trade-off between too few photon events and too many invalid multiphoton occurrences. As shown in Fig.~\ref{fml1}, this balance is achieved near $\mu_t \approx 1$, which minimizes the average synchronization error. Selecting $\mu_t$ carefully around this optimal region is essential for reliable and high-precision quantum synchronization under constrained photon budgets and noisy channel conditions.

\begin{figure}
	\begin{center}
		\includegraphics[width=3.4 in]{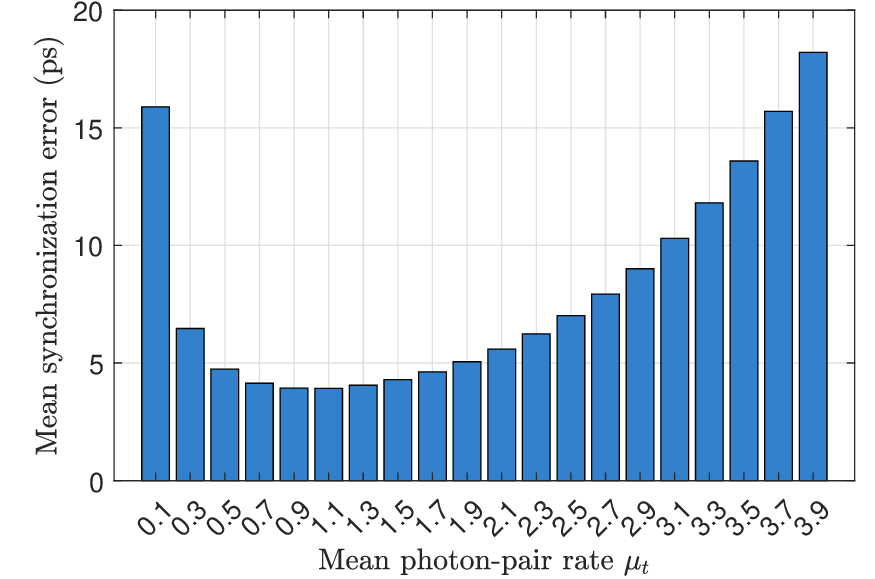}
		\caption{Average synchronization error versus photon-pair generation rate $\mu_t$ for grid size $15 \times 15$ and synchronization duration $T_{\text{seq}} = 1\,\text{ms}$. }
		
		\label{fml1}
	\end{center}
\end{figure}

\textcolor{black}{\subsection{User Mobility, Handover, and Quasi-Static Assumption}}

\textcolor{black}{User mobility is modeled under a quasi-static assumption, which is justified by the short duration of the synchronization window relative to typical indoor motion. For a synchronization interval $T_{\text{sync}}$ on the order of milliseconds (as evidenced in Fig. \ref{gb}), and a maximum indoor walking speed $v \leq 1.5~\mathrm{m/s}$, the resulting displacement satisfies $\Delta d = v T_{\text{sync}} \ll w_z$, where $w_z$ denotes the beam footprint at the receiver plane. Consequently, user motion within a synchronization interval does not significantly affect the photon-reception probability or timing statistics. Mobility therefore manifests as a sequence of quasi-static intervals separated by grid handovers, rather than as continuous time variation. Upon handover, synchronization is re-established using the same procedure without reliance on prior state information, thereby validating the quasi-static assumption adopted in the simulations.}

\textcolor{black}{\subsection{Impact of Spatial Uncertainty and Grid Resolution}}

\textcolor{black}{In this subsection, we evaluate the robustness of the proposed synchronization framework under realistic user-positioning uncertainties. Positioning errors do not affect the photon generation or timing process itself; instead, their impact is purely geometric. Specifically, a mismatch between the estimated and true user location reduces the photon-reception probability by shifting the receiver away from the beam center, thereby decreasing the number of detected photon pairs available for synchronization.}

\textcolor{black}{To quantify this effect, four representative positioning error models are considered while maintaining the same RMS uncertainty $\sigma_p$ for fair comparison. Specifically, we consider: (i) a Gaussian model with independent zero-mean errors (discussed in Section \ref{modeling-sec} as the baseline), $n \sim \mathcal{N}(0,\sigma_p^2 I)$; (ii) a Laplacian model with heavier tails, whose probability density is given by $f(x)=\frac{1}{2b}\exp(-|x|/b)$ with $b=\sigma_p/\sqrt{2}$; (iii) a spatially correlated Gaussian model reflecting correlation
induced by anchor geometry in trilateration-based systems with covariance $\Sigma=\sigma_p^2\begin{bmatrix}1 & \rho & 0\\ \rho & 1 & 0\\ 0 & 0 & 1\end{bmatrix}$; and (iv) a biased model combining Gaussian noise with a uniform offset, i.e., $n_{\text{Uniform}} \sim \mathcal{U}(-0.75\sigma_{p},\, 0.75\sigma_{p})$, to emulate systematic localization errors.  For each case, the photon-reception probability is computed using the beam-based reception model in \eqref{beam_prop_ex}, and the resulting synchronization error is obtained from the Cramér–Rao bound in \eqref{eq:crlb_base}.}

\textcolor{black}{Table~ IV summarizes the resulting performance. The results show that different positioning error statistics primarily affect the photon-reception probability, while the timing statistics of individual detections remain unchanged. Even under the worst-case Laplacian model, the synchronization error increases by only 12\%, requiring approximately 25\% longer integration time to achieve the same precision. This demonstrates that the proposed synchronization method exhibits graceful degradation rather than abrupt failure. Importantly, the observed performance loss arises from reduced photon counts rather than increased timing noise. As a result, synchronization accuracy can be recovered by extending the observation window, confirming that the proposed framework is robust to realistic indoor localization errors and does not rely on idealized positioning assumptions.}

\begin{table}[h]
	\centering
	\label{tab:pos_error}
	\color{black}
	\caption{Synchronization performance under different user-positioning error models}
	\label{tab:noise_distributions}
	\renewcommand{\arraystretch}{1.1}
	\setlength{\tabcolsep}{9pt}
	\footnotesize
	\begin{tabular}{lcccc}
		\hline
		Dist. & $P_{\text{rec}}$ & Sync Err. & Time & Penalty \\
		&                  &     & Factor & (dB) \\
		\hline
		Gaussian   & 0.0317 & 9.2 ps & $1.00\times$ & 0.0 \\
		Laplacian  & 0.0254 & 10.3 ps & $1.25\times$ & 1.0 \\
		Correlated & 0.0280 & 9.8 ps & $1.13\times$ & 0.6 \\
		Biased     & 0.0299 & 9.5 ps & $1.06\times$ & 0.3 \\
		\hline
	\end{tabular}
	
	\vspace{1 mm}
	\footnotesize
	\textit{Simulation parameters}: $w_z = 0.283$~m, $r_a = 0.02$~m, $\sigma_p = 0.2$~m, $\sigma_d = 200$~ps, $\eta_d = 0.6$, $\rho$ = 0.7, and synchronization window of $1$~ms.
\end{table}

\textcolor{black}{\subsection{Impact of Multipath Dispersion and Non-Gaussian Timing Jitter}}

\textcolor{black}{Fig. \ref{fig:new-delay} illustrates the impact of multipath dispersion and non-Gaussian timing jitter on the synchronization accuracy as a function of the synchronization window length. Unlike the idealized LOS case, the synchronization error does not decay to sub 10 ps as the observation time increases. Instead, the error converges to a non-zero floor (i.e., RMS delay), which is a direct consequence of multipath-induced delay spread in the indoor channel. This behavior confirms that, beyond a certain integration duration, additional photon detections do not improve timing accuracy because the dominant impairment is no longer stochastic noise but a deterministic bias introduced by delayed reflected components. The effect is further exacerbated under Laplacian timing jitter, where the heavier tail slows convergence relative to the Gaussian case, even though the mean timing error remains bounded. Importantly, the observed error floor is not a limitation of the proposed synchronization algorithm but a physical consequence of the propagation environment. Since the multipath bias depends on room geometry, surface reflectivity, and transmitter–receiver placement, it can in principle be calibrated and compensated. After such calibration, the residual timing error remains in the picosecond regime, demonstrating that the proposed framework preserves high-precision synchronization performance even under realistic non-LOS and non-Gaussian conditions.}

\begin{figure} [h!]
	\color{black}
	\begin{center}
		\includegraphics[width=3.4 in]{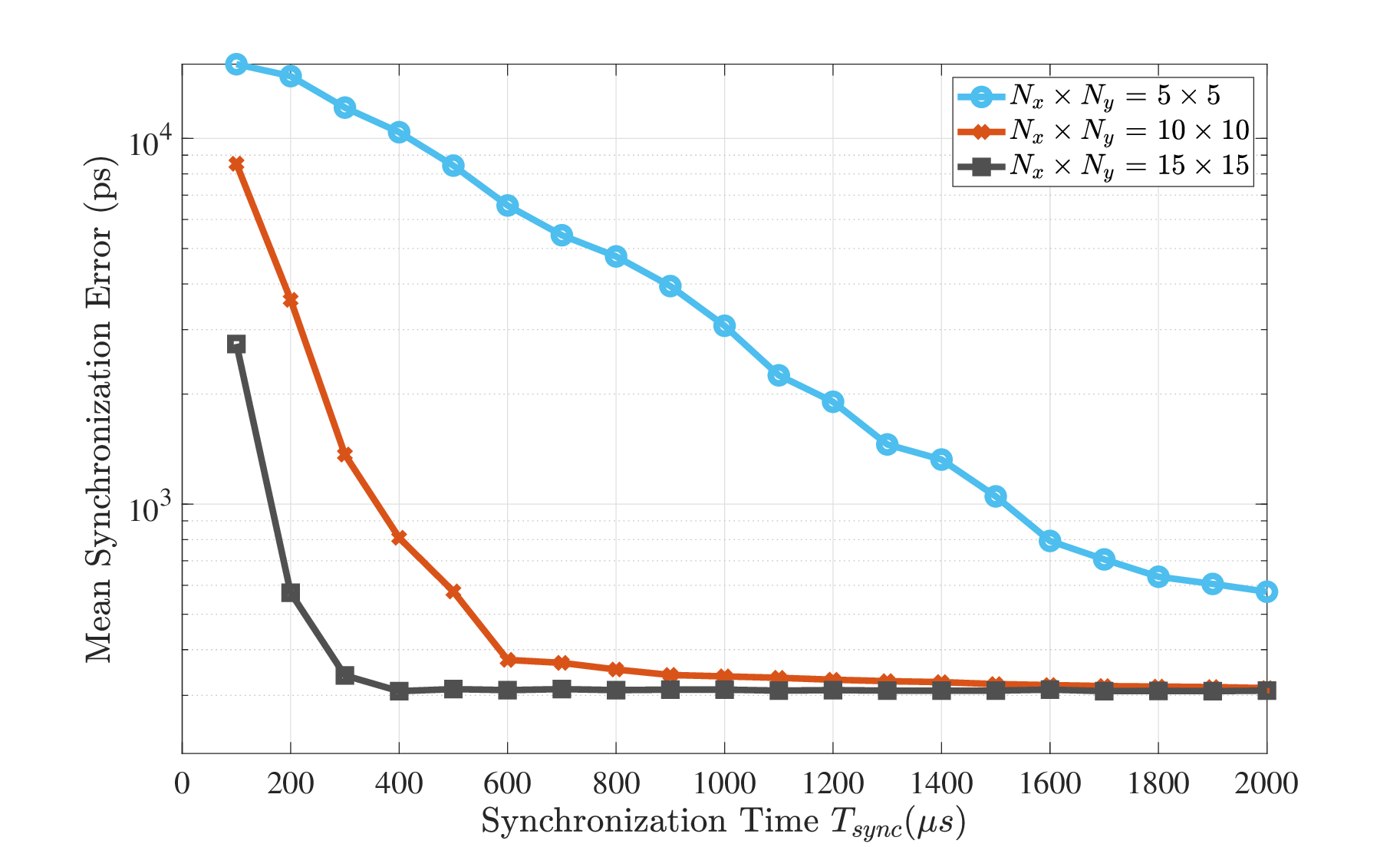}
		\caption{ Average synchronization error as a function of the sequence duration $T_{\text{seq}}$ for several {spatial grid sizes}, evaluated under NLOS channel conditions and non‑Gaussian timing jitter.}
		
		\label{fig:new-delay}
	\end{center}
\end{figure}

\textcolor{black}{\subsection{Computational Complexity and Real-Time Feasibility}}

\textcolor{black}{The computational implications of the proposed synchronization framework are also evaluated to assess its feasibility for real-time implementation. The two-stage synchronization algorithm is intentionally designed to be computationally efficient. The coarse alignment stage relies on sequence correlation, which can be implemented using FFT-based convolution with a computational complexity of $O(N_s \log N_s)$, where $N_s$ denotes the number of time slots. For the largest sequence length considered in this work ($N_s = 2 \times 10^5$), this operation requires only a few milliseconds on contemporary DSP or FPGA platforms, consistent with reported real-time FFT performance \cite{senobari2019super}. The memory footprint is similarly modest, i.e., the binary detection sequence requires approximately $25\,\text{KB}$, while the associated timestamp vector occupies approximately $1.6\,\text{MB}$. Importantly, due to the sparse nature of photon arrivals, only a small fraction of slots contain valid detections, which significantly reduces the effective processing load in practice. As a result, the algorithm is well suited for real-time operation, including fast re-synchronization during user mobility and grid handovers. These results confirm that the proposed synchronization framework is computationally viable for practical indoor optical wireless deployments.}

\section{Conclusion and Future Work}
\label{conclusion-seccc}

\textcolor{black}{In this work, we presented a unified framework for entanglement assisted time synchronization in indoor optical wireless systems, with particular emphasis on grid based architectures that support mobility and spatial reuse. Unlike conventional synchronization approaches that assume static point to point links, the proposed framework explicitly captures the coupling between spatial beam geometry, photon reception statistics, and synchronization accuracy. Particularly, we developed a comprehensive system model that integrates photon pair generation, grid based beam delivery, spatial uncertainty, and realistic detector behaviour. The analysis showed that synchronization performance is fundamentally determined by the number of successfully detected photon pairs, which is shaped by beam alignment and user position uncertainty. Through a Cramér Rao based formulation, we also established a direct quantitative relationship between spatial design parameters and achievable timing precision.}

\textcolor{black}{The impact of practical impairments was further examined through detailed modeling and simulation. Multipath propagation was found to introduce a bounded timing bias rather than instability, while non Gaussian detector jitter primarily affects convergence speed rather than steady state accuracy. These effects define fundamental performance limits but do not undermine the viability of the proposed approach. Even under realistic indoor channel conditions, the framework achieves synchronization accuracy on the order of a few picoseconds. Furthermore, the computational feasibility of the method was demonstrated. The two stage synchronization algorithm supports efficient FFT based implementation with modest memory requirements. Thus, the prposed algorithm is suitable for real time operation. }

\textcolor{black}{Several research directions can be built upon this work. For instance, future efforts may include experimental validation using a hardware testbed, adaptive calibration techniques to mitigate multipath induced bias, and extensions to multi user and multi cell scenarios. In addition, integration with hybrid RF optical architectures and higher layer network protocols can be explored to enable fully deployable quantum enabled indoor communication systems.}

\appendix

\section{}\label{AppA}
Based on \eqref{sq1}, we have
\begin{align}
	\hat{\mathbf{z}}' = \frac{\mathbf{d}^{(i,j)}}{\|\mathbf{d}^{(i,j)}\|}.
\end{align}
To construct the orthonormal basis for the local coordinate system, we choose a reference vector (not parallel to $\hat{\mathbf{z}}'$), e.g.,
\begin{align}
	\mathbf{v}_{\text{ref}} = 
	\begin{cases}
		[0, 0, 1], & \text{if } \hat{\mathbf{z}}' \text{ is not close to } [0, 0, 1] \\
		[1, 0, 0], & \text{otherwise}
	\end{cases}
\end{align}
Then define:
\begin{align}
	\hat{\mathbf{x}}' &= \frac{ \mathbf{v}_{\text{ref}} \times \hat{\mathbf{z}}' }{ \| \mathbf{v}_{\text{ref}} \times \hat{\mathbf{z}}' \| }, \\
	\hat{\mathbf{y}}' &= \hat{\mathbf{z}}' \times \hat{\mathbf{x}}'.
\end{align}
The resulting rotation matrix from global $(x,y,z)$ to beam-aligned $(x',y',z')$ coordinates is:
\begin{align}
	\mathbf{R}^{(i,j)} = \left[ \hat{\mathbf{x}}'^\top \; \hat{\mathbf{y}}'^\top \; \hat{\mathbf{z}}'^\top \right]^\top
\end{align}

Any point $\mathbf{p}$ in the global coordinate system is then transformed into the beam-aligned frame as:
\begin{align}
	\mathbf{p}' = \mathbf{R}^{(i,j)} \left( \mathbf{p} - \mathbf{p}_{\text{tx}} \right)
\end{align}

\bibliographystyle{IEEEtran}
\balance
\bibliography{IEEEabrv,myref}

\end{document}